\documentclass[12pt]{article}
\usepackage{hyperref}
\usepackage{amsmath,amssymb}
\usepackage{graphicx}
\usepackage{color}
\definecolor{myblue}{rgb}{0.14,0.11,0.49}
\definecolor{myred}{rgb}{0.74,0.22,0.15}
\definecolor{mygreen}{rgb}{0.05,0.52,0.42}
\definecolor{myyellow}{rgb}{0.96,0.92,0.13}
\definecolor{myorange}{rgb}{1,0.61,0.36}
\definecolor{mypurple}{rgb}{0.71,0.02,1}
\definecolor{noir}{gray}{0.} % black

\definecolor{htc}{rgb}{1,1,1} % heading text colour
\newcommand{\Mat}[1]{{{\boldsymbol{#1}}}}

\def\be{\begin{equation}}
\def\ee{\end{equation}}
\def\bea{\begin{eqnarray}}
\def\eea{\end{eqnarray}}
\def\bc{\begin{center}}
\def\ec{\end{center}}
\def\bi{\begin{itemize}}
\def\ei{\end{itemize}}
\def\bs{\begin{slide}}
\def\es{\end{slide}}
\def\dd{\mathrm{d}}

\def\noi{\noindent}
\title{On the equations of electrodynamics in a f\mbox{}lat or a curved spacetime and a possible interaction energy}
\author{
Mayeul Arminjon\\
\small\it Univ. Grenoble Alpes, CNRS, Grenoble INP%\footnote{\ Institute of Engineering Univ. Grenoble Alpes}
, 3SR, F-38000 Grenoble, France
} 
\date{}
			     
\begin{document}
%%%%%%%%%%%%%%%%%%%%%%%%%%%%%%%%%%%%%%%%%%%%%%%%%%%%%%%%%%%%%%%%%%%%%%%%%%%%%%%%

\maketitle

\begin{abstract}

\noi We investigate which are the independent equations of continuum electrodynamics and what is their number, beginning with the standard equations used in special and in general relativity. We check by using differential identities that there are as much independent equations as there are unknowns, for the case with given sources as well as for the general case where the motion of the charged medium producing the field is unknown. Then we study that problem in an alternative theory of gravity with a preferred reference frame, in order to constrain an additional, ``interaction" energy tensor that has to be postulated in this theory, and that would be present also outside usual matter. In order that the interaction tensor be Lorentz-invariant in special relativity, it has to depend only on a scalar field $p$. Since the system of electrodynamics of the theory is closed in the absence of the interaction tensor, just one scalar equation more is needed to close it again in the presence of $p$. We add the equation for charge conservation. We derive equations that will allow one to determine the field $p$ in a given weak gravitational field and in a given electromagnetic field.\\

\noi {\bf Keywords:} Maxwell equations, special relativity, general relativity, alternative theory of gravitation, preferred reference frame\\

\noi {\bf PACS:} 04.25.Nx, 04.40.-b, 04.50.-h, 98.58.Ay

%\vspace{80mm}
\end{abstract}

%%%%%%%%%%%%%%%%%%%%%%%%%%%%%%%%%%%%%%%%%%%%%%%%%%%%%%%%%%%%%%%%%%%%%%%%%%%%%%%%
%\tableofcontents
%%%%%%%%%%%%%%%%%%%%%%%%%%%%%%%%%%%%%%%%%%%%%%%%%%%%%%%%%%%%%%%%%%%%%%%%%%%%%%%%

%%%%%%%%%%%%%%%%%%%%%%%%%%%%%%%%%%%%%%%%%%%%%%%%%%%%%%%%%%%%%%%%%%%%%%%%%%%%%%%%
\section{Introduction and summary}\label{Intro}
%%%%%%%%%%%%%%%%%%%%%%%%%%%%%%%%%%%%%%%%%%%%%%%%%%%%%%%%%%%%%%%%%%%%%%%%%%%%%%%%

The usual approach to classical electrodynamics considers, on one hand, a system of point charges, and on the other hand, the electromagnetic (e.m.) field that both is produced by them and acts upon them \cite{L&L,Jackson1998}. The ideal notion of point charge leads to well-known difficulties such as self-force, infinite energy, run-away solutions,... , but actually in the Maxwell equations the charge and current distribution is a continuous one, or at least is preferably so. From the definitions of the ``free" densities of electric charge and current \cite{Jackson1998}, it certainly follows that the distribution should be continuous for the macroscopic Maxwell equations (the ones that contain the displacement field ${\bf D}$ and the magnetizing field ${\bf H}$ in addition to the macroscopic electric field ${\bf E}$ and the macroscopic magnetic field ${\bf B}$). Now, for a linear and isotropic response, the macroscopic equations have the same form as the ``microscopic" Maxwell equations (i.e. the most standard ones, with only the microscopic electric field ${\bf E}$ and the microscopic magnetic field ${\bf B}$) \cite{Jackson1998}. Hence, we may and will study the Maxwell equations without  ${\bf D}$ and without  ${\bf H}$, for a {\it continuous charged medium} and its e.m. field. This applies to a macroscopic, practical situation, if one considers then the particular case of a linear and isotropic medium. It applies also to the general situation at a ``microscopic but still classical" scale, if one considers the total microscopic charge density and the total microscopic current density, without doing any distinction between ``free" and ``bound" charges. However, in the case that the source is not considered given, one has to assume that the velocity of the electric current is the velocity of the charged medium, see Eqs. (\ref{J^mu=rho v^mu}) and (\ref{j=rho v}) below. \\

The initial problem that we were investigating in this work is the development of continuum electrodynamics in an alternative, scalar theory of gravitation with a preferred reference frame or ``ether", in short the scalar ether theory or SET. An alternative extension of Maxwell's second group to the situation with gravitation, consistent with SET, had been proposed in a previous work \cite{A54}. That alternative second group predicted charge non-conservation in a variable gravitational field. However, in a later work \cite{A56}, it has been found that the charge production/destruction thus predicted seems much too high. It has also been found \cite{A56} that the cause for this failure is that in SET one should {\it not} assume that the total energy(-momentum-stress) tensor $\Mat{T}$ (whose $T^{0 0}$ component in the preferred frame is the source of the gravitational field \cite{A35}) is the sum of the energy tensors of the charged medium and the e.m. field: $\Mat{T} \ne \Mat{T}_\mathrm{chg} + \Mat{T}_\mathrm{field}$. The main aim of the present work was therefore to derive explicit equations that should allow one later to calculate the ``interaction energy tensor" 
\be\label{Def T inter}
\Mat{T}_\mathrm{inter} := \Mat{T}- \Mat{T}_\mathrm{chg}- \Mat{T}_\mathrm{field}
\ee
in relevant situations. This is interesting not only for SET in itself, but also because it might be the case that the interaction energy contribute to the ``dark matter" \cite{A56}. To assess $\Mat{T}_\mathrm{inter}$, we must precise the constraints that are imposed on it, hence we must investigate in some detail which are the independent equations of continuum electrodynamics and what is their number. \\

It is instructive and in fact necessary for our purpose to study that problem first for the case of special relativity (SR) and general relativity (GR). This will be the task pursued in {\it Section \ref{Standard}.} We shall show there that the notion of differential identity allows one to explain simply why the eight components of the standard Maxwell equations are needed to determine the six components of ${\bf E}$ and ${\bf B}$, when the 4-current is given; and this same notion allows us to show also that the more complete system obtained by adding the equation of motion of the charged continuum is closed. (For that system the 4-current is an unknown.)\\

The rest of the paper is devoted to the electrodynamics of SET and to the equations for the interaction tensor. In {\it Section \ref{SET_without_T3}} we summarize the situation \cite{A54} when the ``additivity assumption" 
(equivalent to $\Mat{T}_\mathrm{inter}={\bf 0}$ in (\ref{Def T inter})) is made, and, using the same method as in Sect.  \ref{Standard}, we find that also here the equations form a closed system of PDE's. {\it Section \ref{Interact}} brief\mbox{}ly exposes the reasons \cite{A56} that enforce us to abandon this assumption. Then {\it Sect. \ref{Constraints}} studies the constraints that are imposed on $\Mat{T}_\mathrm{inter}$. The demand that it be Lorentz-invariant in SR leads us to the simple form (\ref{T_inter_mixed}), depending on a scalar field $p$.  Just one scalar equation is then lacking in the electrodynamics of SET, due to the introduction of just one new unknown $p$. It follows therefrom that Maxwell's second group cannot be the same in ``SET with $p$" as it is in GR and in the other metric theories of gravitation. In {\it Sect. \ref{ChargeConservn},} we consistently close the system of electrodynamics in ``SET with $p$" by adding the charge conservation. We show that the field $p$ is constant (and arguably zero) for a gravitational field that is constant in the preferred reference frame assumed by that theory. Hence, in particular, $p$ is constant, and arguably zero, in SR with this system of equations. Finally, {\it Sect. \ref{WeakField-p}} establishes the explicit equation that determines the field $p$ in a weak and slowly varying gravitational field, Eq. (\ref{advec_p}), and proposes an integration procedure to get the numerical values of $p$.

%\newpage
%%%%%%%%%%%%%%%%%%%%%%%%%%%%%%%%%%%%%%%%%%%%%%%%%%%%%%%%%%%%%%%%%%%%%%%%%%%%%%%%
\section{Independent equations in standard theory}\label{Standard}
%%%%%%%%%%%%%%%%%%%%%%%%%%%%%%%%%%%%%%%%%%%%%%%%%%%%%%%%%%%%%%%%%%%%%%%%%%%%%%%%

By ``standard theory", we mean the electrodynamics in GR, or possibly in another metric theory of gravity. This includes SR as the case that the spacetime metric is Minkowski's. In any such theory, the dynamical equation verified by the {\it total} energy(-momentum-stress) tensor $\Mat{T}$ of matter and non-gravitational fields is 
\be\label{DT-GR}
T^{\mu \nu }_{\ \, ;\nu}=0;
\ee
and, more generally, there is a rule to go from any equation valid in special relativity in Cartesian coordinates to an equation valid with a general Lorentzian metric in general coordinates: ``comma goes to semicolon" \cite{MTW}, i.e., partial derivatives have to be replaced by covariant derivatives based on the connection associated with the spacetime metric $\Mat{\gamma }$. (This rule becomes ambiguous in cases with derivatives of order larger than one, but one may avoid this in the case of the Maxwell equations: they involve only the e.m. field tensor $\Mat{F}$, not the e.m. 4-potential ${\bf A}$.) That rule does lead to the standard equations for the Maxwell field in a curved spacetime: the first group,
\be\label{Maxwell 1}
F_{\lambda \mu \, ,\,\nu } + F_{\mu \nu \,,\,\lambda 
} + F_{\nu \lambda \,,\,\mu } = F_{\lambda \mu \, ;\,\nu } + F_{\mu \nu \,;\,\lambda 
} + F_{\nu \lambda \,;\,\mu } = 0
\ee 
(the first equality is an identity due to the antisymmetry of $\Mat{F}$ ($F_{\mu \nu }=-F_{\nu \mu }$) and to the symmetry of the metric connection), and the second group,
\footnote{\
In this paper, as in the foregoing \cite{A56}, we are using the SI units and the $(+\ -\ -\ -)$ signature. (In Refs. \cite{L&L, Fock1964, A54}, the Gauss units were used.) Greek indices go from $0$ to $3$, Latin ones from $1$ to $3$. Indices are raised or lowered using the spacetime metric $\Mat{\gamma }$.
}
\be\label{Maxwell 2 GR}
F^{\mu \nu } _{\ \ \,;\nu  }= -\mu_0 J^\mu.
\ee
(Here $J^\mu \ (\mu=0,...,3)$ are the components of the e.m. 4-current ${\bf J}$ and $\mu _0$ is the permeability of free space.) How many independent unknowns do we have in standard theory, and how many independent equations? First, since the first group can be written as
\be\label{M_lambda mu nu}
M_{\lambda \mu \nu } := F_{\lambda \mu \, ;\,\nu } + F_{\mu \nu \,;\,\lambda 
} + F_{\nu \lambda \,;\,\mu }  =0,
\ee
and since the l.h.s. is totally antisymmetric, i.e., $M_{\lambda \mu \nu }=-M_{ \mu \lambda \nu }=-M_{\lambda \nu \mu }$, it is usual to note that (\ref{Maxwell 1}) contains exactly four linearly-independent equations: for example, $M_{0 1 2}=0$, $M_{0 1 3}=0$, $M_{0 2 3}=0$, $M_{1 2 3}=0$. Also, the four equations in the second group (\ref{Maxwell 2 GR}) are linearly independent.

\subsection{Case with given 4-current}\label{J given}

Most often in the discussion of the solutions to the Maxwell equations, the 4-current ${\bf J}$ is considered given. Then we have the eight Maxwell equations (\ref{Maxwell 1})--(\ref{Maxwell 2 GR}) for the six independent unknowns $F_{\mu \nu }\ (0\leq \mu<\nu\leq 3)$. It is nevertheless well known that those eight equations are needed, i.e., one cannot consistently remove two of them from the whole. In particular, one cannot consistently remove the two divergence equations  $\mathrm{div}\,{\bf B}=0$ and $\mathrm{div}\, {\bf E}=\rho _\mathrm{el}/\epsilon _0$ from the usual three-vector form of the f\mbox{}lat-spacetime Maxwell equations \cite{Jiang1996,Zhou2006,Liu2017}. 
\footnote{ \label{F vs E-B}
Here $\rho _\mathrm{el}$ is the volume density of the electric charge and $\epsilon _0$ is the permittivity of the vacuum. With the SI units and the $(+\ -\ -\ -)$ signature, the relation between the field tensor $\Mat{F}$ and the electric and magnetic fields ${\bf E}$ and ${\bf B}$ in Cartesian coordinates in a Minkowski spacetime is given on the Wikipedia page ``Electromagnetic tensor", or by Eq. (47) of Ref. \cite{A56}.
}
Regarding this point, it is observed \cite{Jiang1996} that the Maxwell equations form two pairs of ``div-curl" systems of equations, a div-curl system consisting of four scalar equations for three scalar unknowns $u^i\ (i=1,2,3)$:
\be\label{div-curl}
\mathrm{div}\,{\bf u} =f  ,\quad \mathrm{rot}\,{\bf u} ={\bf s}.
\ee
Starting from this observation, the following has been shown in a detailed mathematical work \cite{Jiang1996}: (i) There is a theorem stating uniqueness of the solutions ${\bf u}$ to the div-curl system, modulo suitable boundary conditions, and provided the data ${\bf s}$ verifies the compatibility condition
\be\label{div rot}
\mathrm{div}\,{\bf s} = 0
\ee
(which follows from applying the differential identity $\mathrm{div}\,\mathrm{rot}\,{\bf u} \equiv 0$ to Eq. (\ref{div-curl})$_2$). (ii) Removing the divergence equation in one div-curl system leads to a system whose solutions are not necessarily solutions of that div-curl system.\\

However, in our opinion, the fact that the div-curl system, or respectively the Maxwell equations, are not overdetermined, is more easily understood %--- without necessarily having recourse to a detailed mathematical work like \cite{Jiang1996} --- 
from the observation that these systems of PDE's verify definite {\it differential identities}. This notion (without its name) is recognized by Liu \cite{Liu2017} as relevant, see his Definition II, but unfortunately he does not indicate {\it which} differential identities do apply to the div-curl system (\ref{div-curl}) or respectively to the Maxwell equations. The identity $\mathrm{div}\,\mathrm{rot}\,{\bf u} \equiv 0$, which he mentions, is not a differential identity {\it of the div-curl system (\ref{div-curl}),} because $\mathrm{rot}\,{\bf u} = {\bf 0}$ does not belong to the equations of that system (unless ${\bf s}\equiv {\bf 0}$). Also, Eq. (\ref{div rot}) is obviously not a differential identity, i.e., a partial differential equation that is satisfied by any regular vector function ${\bf s}$. What happens is that, {\it if} Eq. (\ref{div rot}) is satisfied by some particular function ${\bf s}$, then the following is indeed a differential identity, i.e., it is valid for any regular vector function ${\bf u}$:
\be\label{div_rot u_s}
\mathrm{div}\,(\mathrm{rot}\,{\bf u} -{\bf s})\equiv 0.
\ee
Moreover, this is a differential identity {\it of the system (\ref{div-curl}),} i.e., it has the form 
\be\label{Diff Id System}
\sum _{k=1} ^{n} \mathcal{O}_k \mathcal{P}_k {\bf u} \equiv 0,
\ee
where $\mathcal{P}_k {\bf u} =0 $ ($k=1,...,n$, thus $n=4$ for the system (\ref{div-curl})) are the scalar PDE's of the system, and where $\mathcal{O}_k$ are scalar %linear 
differential operators of the first order, or the zero operator. Indeed some among the $\mathcal{O}_k$ 's, but not all, can be the zero operator, meaning that the corresponding equation $\mathcal{P}_k {\bf u} =0$ is not involved in the differential identity. (Also, ${\bf u}$ designates in general the list of unknown functions in the system.) We insist that, by definition, a differential identity like (\ref{Diff Id System}) has to apply whether or not any of the equations of the system ($\mathcal{P}_k {\bf u} =0$) is satisfied. Thus, in view of the scalar identity (\ref{div_rot u_s}), it is clear that the system (\ref{div-curl}) has only three independent equations instead of four. But since it is a differential identity, not an algebraic one, and since that identity involves not merely one but several equations among the scalar PDE's of the system, it is quite clear also that we can't remove any of the scalar PDE's of the system without altering it. In particular, note that (under the validity of the compatibility condition (\ref{div rot})), the differential identity (\ref{div_rot u_s}) is already a differential identity {\it of the ``curl system" (\ref{div-curl})$_2$ alone,} which is hence an {\it underdetermined} system: $3-1=2$ independent equations for 3 unknowns. It thus becomes obvious that, if one removes the divergence equation (\ref{div-curl})$_1$, one will indeed be able to find ``spurious" solutions, i.e., undesired ones. \\

Consider now the f\mbox{}lat-spacetime Maxwell system {\it in vacuo:} Eqs. (\ref{Maxwell 2 GR})--(\ref{M_lambda mu nu}) with commas instead of semicolons (in Cartesian coordinates for the Minkowski metric), and with $J^\mu=0$. As noted by Das \cite{Das1996}, that system has the two differential identities
\be\label{Das_SR}
S:= \varepsilon _{\mu \nu \rho \sigma } M_{\mu \nu \rho ,\sigma } \equiv 0
\ee
(where $\varepsilon _{\mu \nu \rho \sigma } $ is the signature of the permutation $(\mu \nu \rho \sigma)$ of $\{0,...,3\}$), and
\be\label{divdivF=0_SR}
F^{\mu \nu } _{\ \ \,,\nu,\mu   }\equiv 0,
\ee
which both result from the antisymmetry of the tensor $\Mat{F}$ (after a short algebra, for (\ref{Das_SR})). The first identity is a differential identity of the first Maxwell group (\ref{M_lambda mu nu}), the second one is a differential identity of the second Maxwell group {\it in vacuo.} Thus there are six independent equations for the six unknowns. 
\\

Finally, consider the Maxwell system in a general spacetime (\ref{Maxwell 2 GR})--(\ref{M_lambda mu nu}). In a general coordinate system in a general Lorentzian spacetime, we have similarly with (\ref{Das_SR}) and (\ref{divdivF=0_SR}) the two independent differential identities
\be\label{Das_GR}
e^{\mu \nu \rho \sigma } M_{\mu \nu \rho ;\sigma } \equiv 0
\ee
(where $e _{\mu \nu \rho \sigma } $ is the usual totally antisymmetric tensor that coincides with $\varepsilon _{\mu \nu \rho \sigma }$ in coordinates such that the natural basis is direct and that $\gamma :=\mathrm{det}(\gamma _{\mu \nu })=-1$), and
\be\label{divdivF=0}
F^{\mu \nu } _{\ \ \,;\nu;\mu   }\equiv 0.
\ee
These identities indeed apply (thus whether or not Eqs. (\ref{Maxwell 2 GR}) or (\ref{M_lambda mu nu}) are verified), because the l.h.s. of either (\ref{Das_GR}) or (\ref{divdivF=0}) is a manifestly invariant scalar, and it is zero (as for (\ref{Das_SR}) or (\ref{divdivF=0_SR})) in coordinates such that, at the event considered, the Christoffel symbols vanish and the matrix $(\gamma _{\mu \nu })=(\eta _{\mu \nu }):=\mathrm{diag}(1,-1,-1,-1)$. The first one, Eq. (\ref{Das_GR}), is clearly a differential identity of the first Maxwell group (\ref{M_lambda mu nu}). The identity (\ref{divdivF=0}) implies that the second Maxwell group (\ref{Maxwell 2 GR}) has charge conservation as a compatibility condition:
\be\label{ChargeConserv}
J^\mu _{\ \, ;\mu }=0.
\ee
Hence, it is here as with the div-curl system (\ref{div-curl}): if the compatibility condition (\ref{ChargeConserv}) is satisfied, as it should, then we get from (\ref{divdivF=0}) the following differential identity of the second Maxwell group (\ref{Maxwell 2 GR}):
\be\label{div_F+J=0}
S' := \left (F^{\mu \nu } _{\ \ \,;\nu  }+\mu_0 J^\mu \right)_{;\mu}\equiv 0.
\ee
Thus, here also, we have six independent equations for the six unknowns.\\

It may be useful to show how the identities (\ref{Das_GR}) and (\ref{div_F+J=0}) appear with the usual 3-vector form of the Maxwell equations for a f\mbox{}lat spacetime. Using Eqs. (24.13) and (24.14) of Fock \cite{Fock1964}, it is easy to check that the l.h.s. of the identity (\ref{Das_SR}) can be rewritten as
\be\label{Das-3-vector}
S = 6\left[ \mathrm{div} \left (\mathrm{rot}\, {\bf E} + \frac{\partial {\bf B}}{\partial t} \right )-\frac{\partial}{\partial t} \left( \mathrm{div}\, {\bf B} \right)\right ].
\ee
In a f\mbox{}lat spacetime, the identity $S\equiv 0$ is thus as well a differential identity of the usual 3-vector form (e.g. Eqs. (1) and (4) in Ref. \cite{Liu2017}) of Maxwell's first group. Using $\mu _0\epsilon _0=1/c^2$, it is also easy to check that, in a f\mbox{}lat spacetime, the identity (\ref{div_F+J=0}) rewrites as the following differential identity of the usual form of Maxwell's second group:
\be\label{div_F+J=0-3-vector}
-S' = \mathrm{div}\left [\mathrm{rot}\, {\bf B}-\mu _0 \left({\bf j}+\epsilon _0 \frac{\partial {\bf E}}{\partial t} \right ) \right ] + \mu _0\epsilon _0 \frac{\partial }{\partial t} \left ( \mathrm{div}\, {\bf E} - \frac{\rho_\mathrm{el}}{\epsilon _0}\right ) \equiv 0.
\ee
This identity is satisfied provided that the integrability condition of Maxwell's second group: Eq. (\ref{ChargeConserv}) or equivalently $\frac{\partial \rho_\mathrm{el} }{\partial t} + \mathrm{div}\, {\bf j}=0$, is satisfied (here ${\bf j}$ is the 3-vector with components $j^i:= J^i$). Of course, as for $S\equiv 0$, the identity (\ref{div_F+J=0-3-vector}) applies whether or not any of the Maxwell equations is satisfied.

\subsection{Complete system for a deformable charged medium}

Often, in practice, the 4-current density  ${\bf J}$ can be considered given (at least as an approximation), as we just envisaged. This is not the general case, however. A continuous charged medium is subjected to the Lorentz 4-force (density)
\be\label{Lorentz 4-force density}
f^\mu  := F^\mu _{\ \, \nu }\,J^\nu,
\ee
which tends to modify the (3-)velocity field ${\bf v}$ of the charged medium (especially if the latter is deformable, as is the case e.g. in magnetohydrodynamics), and hence also to modify the current ${\bf J}$ as well. Indeed, in any coordinates $x^\mu$, the components of the latter 4-vector are defined as (\cite{L&L}, Eq. (90.3)):
\be\label{J^mu=rho v^mu}
J^{\, \mu } := \frac{\rho_\mathrm{el}}{\beta} \frac{\dd x^\mu }{\dd t}, 
\ee
where 
\be\label{beta}
\beta :=  \sqrt{\gamma _{00}},  
\ee
and $t :=x^0/c$. The velocity field ${\bf v}$ (whose value at an event $X$ is the velocity of that infinitesimal volume element of the continuous medium which is at $X$)  is measured with the local standards, and has thus components $v^i=\dd x^i/\dd t_{\bf x}$ where $t_{\mathbf{x}}$ is the synchronized local time in the reference f\mbox{}luid $\mathcal{F}$ that is considered \cite{L&L, Cattaneo1958,B39}. (Note that the data of one coordinate system $(x^\mu)$ automatically defines a unique reference f\mbox{}luid, whose reference world lines are the lines ${\bf x}:=(x^i)=\mathrm{Constant}$ \cite{B39}.) Assuming that $\mathcal{F}$ admits adapted coordinates that verify the synchronization condition 
\be\label{gamma_0i=0}
\gamma _{0 i}=0
\ee
(and using any such coordinates), we have
\be\label{dt_x}
\frac{\dd t_{\bf x}}{\dd t} = \beta (t,{\bf x}),
\ee
so that $J^i=\rho _\mathrm{el} v^i$, or 
\be\label{j=rho v}
{\bf j} = \rho _\mathrm{el} {\bf v}
\ee
for the 3-current, a spatial vector field in the reference f\mbox{}luid $\mathcal{F}$. Moreover, a deformable charged medium has state parameters, at least the proper rest-mass density field $\rho^\ast $ or (equivalently for a barotropic perfect f\mbox{}luid) the pressure field $P$.\\

 In the general case, we thus add to the six field unknowns $F_{\mu \nu }\ (0\leq \mu<\nu\leq 3)$, the field unknowns $\rho _\mathrm{el}$ and ${\bf v}$, plus the other state parameters, say merely the field $\rho^\ast $ for the simplicity of discussion --- thus we have five unknowns more, that is $6+5=11$ unknowns.  The equation we have in addition to the Maxwell equations (\ref{Maxwell 2 GR})--(\ref{M_lambda mu nu}) is the dynamical equation of GR for the charged medium subjected to the 4-force (\ref{Lorentz 4-force density}):
\be\label{Dyn_T-chg-GR}
 T_{\mathrm{chg}\ \,;\nu}^{\mu  \nu}= F^\mu_{\ \ \lambda }\,J^\lambda,
\ee  
where $\Mat{ T}_{\mathrm{chg}}$ is the energy tensor of the deformable charged medium, depending on ${\bf v}$ and $\rho^\ast $. Equation (\ref{Dyn_T-chg-GR}) extends a standard equation valid in special relativity (e.g. \cite{L&L}, Eq. (33.9)), by using the ``comma goes to semicolon" rule. It can also be derived from the dynamical equation for the total energy tensor (\ref{DT-GR}), Maxwell's second group (\ref{Maxwell 2 GR}), and the assumption of additivity of the energy tensors:
\be\label{T=Tcharges+Tfield}
\Mat{T} = \Mat{T}_\mathrm{chg} + \Mat{T}_\mathrm{field}, 
\ee
with $\Mat{T}_\mathrm{field}$ the energy tensor of the e.m. field \cite{L&L, Fock1964}:
\be\label{T field}
T_\mathrm{field}^{\, \mu \nu } := \left (- F^\mu_{\ \ \lambda } F^{\, 
\nu \lambda } + \frac{1}{4}\gamma^{\, \mu \nu } F_{\lambda 
\rho } F^{\lambda \rho \, } \right)/\mu_0 . 
\ee
Indeed, by using the identity \cite{A54,A56}
\be\label{Eq T-field-2}
\mu_0 T_{\mathrm{field}\ \,;\nu}^{\mu  \nu} \equiv  -F^\mu_{\ \ \lambda }\,F^{\nu \lambda }_{\ \ ;\nu },
\ee
one deduces from the second group (\ref{Maxwell 2 GR}):
\be\label{Eq T-field-4}
 T_{\mathrm{field}\ \,;\nu}^{\mu  \nu}= -F^\mu_{\ \ \lambda }\,J^\lambda.
\ee
Equations (\ref{DT-GR}) and (\ref{T=Tcharges+Tfield}) then imply immediately (\ref{Dyn_T-chg-GR}). Thus, we have five unknowns more than in the case with given 4-current, and merely the four equations (\ref{Dyn_T-chg-GR}) more, in addition to (\ref{Maxwell 2 GR}) and (\ref{M_lambda mu nu}). 
\footnote{\
At least if one assumes a perfect f\mbox{}luid, the mass conservation can be exactly {\it deduced} from (\ref{Dyn_T-chg-GR}), in nearly the same way as one deduces it from (\ref{DT-GR}) \cite{Chandrasekhar1969} in the case without the e.m. field and current (in both cases, mass conservation applies iff the f\mbox{}luid is isentropic). Hence it is not an additional equation.
}
However, since in the system [(\ref{Maxwell 2 GR})--(\ref{M_lambda mu nu}), (\ref{Dyn_T-chg-GR})] the current ${\bf J}$ is unknown as well as is the e.m. field $\Mat{F}$, Eq. (\ref{div_F+J=0}) is not a differential identity of that system, even though it does have the form (\ref{Diff Id System}) (with ${\bf u}$ denoting the set of all unknowns of the system: now ${\bf J}$, $\Mat{F}$, and $\rho ^\ast $, or equivalently $\rho _\mathrm{el}$, ${\bf v}$, $\Mat{F}$, and $\rho^\ast $). This is because (\ref{div_F+J=0}), as well as the charge conservation (\ref{ChargeConserv}), apply merely on the {\it solution space} of the system. (Both indeed apply there, due to Maxwell's second group (\ref{Maxwell 2 GR}) and to the identity (\ref{divdivF=0}).) Whereas, a differential identity of the system has to be valid for {\it any} regular field unknowns ${\bf J}$, $\Mat{F}$, and $\rho ^\ast $.  Note also that the identity (\ref{divdivF=0}) is not a differential identity of the system either, because it does not have the form (\ref{Diff Id System}). Thus, for the system [(\ref{Maxwell 2 GR})--(\ref{M_lambda mu nu}), (\ref{Dyn_T-chg-GR})], we have merely the identity (\ref{Das_GR}) that counts as a dependence relation. So we have $12-1=11$ equations for the $11$ unknowns.\\

Note that the foregoing applies in SR as well as in a metric theory of gravitation. However, in the case with gravitation, strictly speaking, the metric field $\Mat{\gamma }$ should not be considered given, e.g. because the e.m. energy tensor does contribute to the total energy tensor and hence inf\mbox{}luences the metric nonlinearly through the Einstein equations (considering GR for definiteness) --- even if this inf\mbox{}luence is very small in usual conditions, e.g. in the solar system. If we account for this, we thus add the ten unknowns $\gamma _{\mu \nu }\ (0\leq \mu \leq \nu \leq 3)$. And we add the ten Einstein equations (of which only six count as independent due to the four differential identities of Bianchi), plus the four equations of the gauge condition which is selected.

%%%%%%%%%%%%%%%%%%%%%%%%%%%%%%%%%%%%%%%%%%%%%%%%%%%%%%%%%%%%%%%%%%%%%%%%%%%%%%%%
\section{Independent equations in SET without interaction tensor}\label{SET_without_T3}
%%%%%%%%%%%%%%%%%%%%%%%%%%%%%%%%%%%%%%%%%%%%%%%%%%%%%%%%%%%%%%%%%%%%%%%%%%%%%%%%

In SET, motion is governed by an extension of the special-relativistic form of Newton's second law to a curved spacetime, written in the preferred reference f\mbox{}luid $\mathcal{E}$ assumed by the theory. This extension, which is formulated primarily for a test particle \cite{A16}, can be applied to each particle 
of a dust, that is an ideal continuum made of a myriad of coherently moving test particles. This leads \cite{A54} to the following dynamical equation in the presence of a field of external (3-)force having volume density ${\bf f}$ with components $f^i\ (i=1,2,3)$:
\be\label{Eq T-f}
T_\mathrm{medium \ \,;\nu}^{\mu \nu} =b^\mu (\Mat{T}_\mathrm{medium})+f^\mu ,
\qquad f^0 := \frac{{\bf f.v}}{c\beta},
\ee
where $\beta $ is defined in Eq. (\ref{beta}), ${\bf v}$ is the velocity field (with the local time, see after that equation), $\Mat{T}_\mathrm{medium}$ is the energy-momentum tensor of the continuous medium, and
\be\label{b^mu}
b^0(\Mat{T}) := \frac{1}{2}\,\gamma^{00}\,g_{ij,0}\,T^{ij},
\quad b^i(\Mat{T}) := \frac{1}{2}\,g^{ij}\,g_{jk,0}\,T^{0k}.
\ee
Here $g_{ij}$ are the components of the spatial metric tensor $\Mat{g}=\Mat{g}_\mathcal{E}$ associated with the spacetime metric $\Mat{\gamma }$ in the reference f\mbox{}luid $\mathcal{E}$ \cite{L&L, Moller1952, A54}. Equation (\ref{Eq T-f}) is then assumed to be valid for any continuous medium \cite{A54}, provided a velocity field can be unambiguously be defined for that medium. In the case that the continuous medium is a charged medium and ${\bf f}$ is the Lorentz (3-)force, the logic of the theory leads one to define its components $f^i$ precisely as the spatial components $\mu =i$ in Eq. (\ref{Lorentz 4-force density}) above \cite{A54}. Moreover, the $f^0$ component  is thus {\it derived} (at least for a dust) to be given by (\ref{Eq T-f})$_2$ here, and this turns out to be also equal to the $\mu =0$ component in Eq. (\ref{Lorentz 4-force density}) \cite{A54} --- although the dynamics is different from that of GR. So for a charged medium the dynamical equation writes:
\be\label{Dyn_T-chg-SET}
T^{\mu  \nu}_{\mathrm{chg}\ \ ;\nu} =b^\mu (\Mat{T}_\mathrm{chg})+F^\mu_{\ \, \nu }\,J^\nu.
\ee
The dynamical equation for the total energy tensor $\Mat{T}$ is also obtained by induction from what is got for a dust, this time without any non-gravitational external force, and is thus \cite{A20}: 
\be\label{DT-SET}
T^{\mu  \nu}_{\ \ ;\nu} =b^\mu (\Mat{T}).
\ee

\vspace{3mm}
It seems natural at first sight to assume the additivity of the energy tensors, Eq. (\ref{T=Tcharges+Tfield}). It turns out that, together with the dynamical equations for the charged medium and the total energy tensor, Eqs. (\ref{Dyn_T-chg-SET}) and (\ref{DT-SET}), this determines Maxwell's second group. This is already true for GR, as one can see by reverting the line of reasoning in Eqs. (\ref{Eq T-field-2}) and (\ref{Eq T-field-4}) above: starting from the dynamical equations for the charged medium and the total energy tensor in GR, Eqs. (\ref{Dyn_T-chg-GR}) and (\ref{DT-GR}), and using the additivity (\ref{T=Tcharges+Tfield}), we obtain immediately Eq. (\ref{Eq T-field-4}). Equating with the identity (\ref{Eq T-field-2}) gives us
\be\label{Eq T-field-5}
 -F^\mu_{\ \ \lambda }\,F^{\nu \lambda }_{\ \ ;\nu } = -\mu _0\,F^\mu_{\ \ \lambda }\,J^\lambda,
\ee
which, at least for the (generic) case of an invertible matrix $(F^\mu_{\ \ \lambda })$, is equivalent to the second group of GR, Eq. (\ref{Maxwell 2 GR}). For SET, this same argument, with Eqs. (\ref{Dyn_T-chg-SET}) and (\ref{DT-SET}) replacing Eqs. (\ref{Dyn_T-chg-GR}) and (\ref{DT-GR}) of GR, leads first to 
\be\label{Eq T-field}
T^{\mu  \nu}_{\mathrm{field}\ \ ;\nu} =b^\mu ({\bf T_\mathrm{field}})-F^\mu_{\ \, \nu }\,J^\nu,
\ee
from which one gets by (\ref{Eq T-field-2}) \cite{A54}:
\be\label{Eq T-field-3}
F^\mu_{\ \ \lambda }\,F^{\lambda \nu }_{\ \ \,;\nu }= \mu_0 \left[ b^\mu \left ({\bf T}_\mathrm{field} \right)-F^\mu_{\ \ \lambda }\,J^\lambda \right ],
\ee
which is the second group got in SET when one assumes the additivity (\ref{T=Tcharges+Tfield}).\\

The foregoing implies that, for SET with the additivity assumption (\ref{T=Tcharges+Tfield}), we can take as the system of equations governing the motion of a charged medium and its e.m. field: Maxwell's first group (\ref{M_lambda mu nu}), plus the dynamical equations for the charged medium and the total energy tensor, Eqs. (\ref{Dyn_T-chg-SET}) and (\ref{DT-SET})  --- or, equivalently, Maxwell's first group (\ref{M_lambda mu nu}), plus Eqs. (\ref{Dyn_T-chg-SET}) and (\ref{Eq T-field-3}). The unknown fields are the same as for GR: $F_{\mu \nu }\ (0\leq \mu<\nu\leq 3)$, $\rho _\mathrm{el}$ and ${\bf v}$, and $\rho^\ast $, or equivalently $F_{\mu \nu }\ (0\leq \mu<\nu\leq 3)$, ${\bf J}$, and $\rho^\ast $ --- in any case 11 unknowns. And as in GR, we have merely the identity (\ref{Das_GR}) that counts as a dependence relation. So again we have $12-1=11$ equations for the $11$ unknowns. If we account for the fact that the gravitational field is not given, we add the scalar gravitational field unknown, $\psi :=-\mathrm{Log}\,\beta $, and we add the scalar f\mbox{}lat-spacetime wave equation obeyed by $\psi $ according to that theory \cite{A35}.

%%%%%%%%%%%%%%%%%%%%%%%%%%%%%%%%%%%%%%%%%%%%%%%%%%%%%%%%%%%%%%%%%%%%%%%%%%%%%%%%
\section{Interaction energy tensor in SET}\label{Interact}
%%%%%%%%%%%%%%%%%%%%%%%%%%%%%%%%%%%%%%%%%%%%%%%%%%%%%%%%%%%%%%%%%%%%%%%%%%%%%%%%

There are three reasons \cite{A56} why it turns out to be not satisfying to close the electrodynamics for SET by stating the additivity assumption (\ref{T=Tcharges+Tfield}):\\

({\bf i}) The version (\ref{Eq T-field-3}) of Maxwell's second group leads to the non-conservation of charge in a variable gravitational field. (This is seen by using the identity (\ref{divdivF=0}) \cite{A54}.) It happens moreover that the amounts of electric charge which are thus predicted to be produced or destroyed in a realistic e.m. and gravitational field, seem much too high to be a tenable prediction.\\

({\bf ii}) In general, the energy tensors of the charged medium and the e.m. field are both non-zero. This means that we are in the presence of a mixture. According to the standard theory of mixtures (which has been developed for non-relativistic physics), the effective energy tensor of the mixture is not the sum of the energy tensors of its constituents \cite{Muller1968}. This is not compatible with the additivity assumption (\ref{T=Tcharges+Tfield}).\\

({\bf iii}) Equation (\ref{Eq T-field}) means that the e.m. field continuum verifies the dynamical equation (\ref{Eq T-f}) for a continuous medium with an external force field ${\bf f}_\mathrm{field}$ acting on it, with, specifically:
\be\label{f_field}
f^\mu _\mathrm{field}= -f^\mu  := -f^\mu _\mathrm{chg}:=-F^\mu  _{\ \,\nu } J^\nu.
\ee
Using the spatial part of this ($\mu =1,2,3$), and noting respectively ${\bf v}_\mathrm{field}$ and ${\bf v}_\mathrm{chg}$ the velocity fields of the field continuum and the charged continuum (assuming ${\bf v}_\mathrm{field}$ is well defined), the ``time" part rewrites as \cite{A56}:
\be\label{f.(v_field-v_charges)=0}
{\bf f}_\mathrm{chg}{\bf .}\left({\bf v}_\mathrm{field} -{\bf v}_\mathrm{chg}\right)=0.
\ee
However%A first difficulty with this is that
, for a general e.m. field, it is not easy to tell how to define its velocity field ${\bf v}_\mathrm{field}$, be it experimentally or in terms of the energy tensor $\Mat{T}_\mathrm{field}$. %The second difficulty is that 
Anyway, Eq. (\ref{f.(v_field-v_charges)=0}) looks problematic for a general e.m. field. For a ``null" field (i.e., such that the classical invariants are both zero), the velocity field can be defined naturally in terms of the energy tensor $\Mat{T}_\mathrm{field}$ \cite{A56}, and it has modulus $v_\mathrm{field}:=(\Mat{g}({\bf v}_\mathrm{field},{\bf v})_\mathrm{field})^{1/2}=c$. Furthermore, one can then prove \cite{A56} that indeed (\ref{f.(v_field-v_charges)=0}) is satisfied (although usually $v_\mathrm{chg}\ll c$), but this proof depends heavily on the fact the e.m. field is a null field.\\

\noi Due to these reasons, we have to abandon the additivity assumption (\ref{T=Tcharges+Tfield}). This means that the total tensor $\Mat{T}$ verifying Eq. (\ref{DT-SET}) involves another part, say $\Mat{T}_\mathrm{inter}$:
\be\label{T_with_interact}
\Mat{T} = \Mat{T}_\mathrm{chg} + \Mat{T}_\mathrm{field} + \Mat{T}_\mathrm{inter}.
\ee
While introducing that ``interaction tensor" $\Mat{T}_\mathrm{inter}$, we almost necessarily add new unknowns --- unless all state parameters for $\Mat{T}_\mathrm{inter}$ would be extracted from those of $\Mat{T}_1:=\Mat{T}_\mathrm{chg}$ and $\Mat{T}_2:=\Mat{T}_\mathrm{field}$.
\footnote{\
For instance, the standard theory of mixtures makes a definite suggestion to define $\Mat{T}$ in terms of $\Mat{T}_1$, $\Mat{T}_2$, and the velocity fields ${\bf v}_1$ and ${\bf v}_2$: Eq. (2.13) in Ref. \cite{Muller1968}. This suggestion can be extended naturally to SET, by starting from Eq. (39) in Ref. \cite{A54} (valid for a dust), but this definition for $\Mat{T}$ does not verify Eq. (\ref{DT-SET}). Anyway, an effective energy tensor defined in such a way would not reduce to the sum (\ref{T=Tcharges+Tfield}) in special relativity, as is required (see Sect. \ref{Constraints}).
}
Hence, the system [(\ref{M_lambda mu nu}), (\ref{Dyn_T-chg-SET}), (\ref{DT-SET})] made of Maxwell's first group and the dynamical equations for the charged medium and for the total energy tensor, cannot be closed any more. So we have to add at least one equation.

%%%%%%%%%%%%%%%%%%%%%%%%%%%%%%%%%%%%%%%%%%%%%%%%%%%%%%%%%%%%%%%%%%%%%%%%%%%%%%%%
\section{Constraints on the interaction tensor}\label{Constraints}
%%%%%%%%%%%%%%%%%%%%%%%%%%%%%%%%%%%%%%%%%%%%%%%%%%%%%%%%%%%%%%%%%%%%%%%%%%%%%%%%

The effect of the gravitational field on the e.m. field can usually be neglected, e.g. because the gravitational force on a charged particle is extremely small as compared with the Lorentz force. When this approximation is done, the e.m. field is as in special relativity (SR), i.e. in a f\mbox{}lat Minkowski spacetime. There is a massive experimental evidence for the predictions deduced from the usual (f\mbox{}lat-spacetime) Maxwell equations. Therefore, we ask that the electrodynamics of SET should exactly reduce to that of SR in the absence of gravitation, i.e., when the scalar gravitational field $\beta \equiv 1$ so that, accordingly \cite{A35}, the physical spacetime metric $\Mat{\gamma }$ is the Minkowski metric. (See Eq. (\ref{gamma}) below.) %When the spacetime metric is Minkowski's, in fact already for a constant gravitational field in SET ($\beta _{,0}=0$), so that $g_{ij,0}=0$, the dynamical equation for the charged medium in SET (\ref{Dyn_T-chg-SET}) reduces to the equation valid in GR (and in SR), Eq. (\ref{Dyn_T-chg-GR}); 
In special relativity, the dynamical equation for the charged medium valid in GR, Eq. (\ref{Dyn_T-chg-GR}), can be derived directly from the Lorentz 4-force (e.g. \cite{L&L}, Eq. (33.9)), while the dynamical equation for the e.m. field valid in GR, Eq. (\ref{Eq T-field-4}), is derived essentially as it was derived hereabove for GR, i.e. from the identity (\ref{Eq T-field-2}) and Maxwell's second group (\ref{Maxwell 2 GR}) (though in a simpler way for SR, with commas instead of semicolons in Cartesian coordinates; see e.g. Eq. (33.7) in Ref. \cite{L&L}). Hence we have in SR as well as in GR:
\be\label{divT_chg+field-GR}
 T_{\mathrm{chg}\ \,;\nu}^{\mu  \nu}+ T_{\mathrm{field}\ \,;\nu}^{\mu  \nu} =0.
\ee  
Again for SR and GR, this is compatible with the additivity assumption (\ref{T=Tcharges+Tfield}): of course, the latter plus the dynamical equation (\ref{DT-GR}) for the total energy tensor imply (\ref{divT_chg+field-GR}). Conversely, if we start from the fully general decomposition (\ref{T_with_interact}) of the total tensor, then under the validity of Eq. (\ref{DT-GR}), we get that (\ref{divT_chg+field-GR}) is equivalent to
\be\label{divT_int=0}
T_{\mathrm{inter}\ \,;\nu}^{\mu  \nu} =0.
\ee
We note that Eq. (\ref{DT-GR}) does apply in SET for a constant gravitational field: the latter means $\beta _{,0}=0$, hence $g_{ij,0}=0$ in (\ref{b^mu}) (see Eq. (\ref{gamma}) below), whence $b^\mu=0$ in (\ref{DT-SET}). Hence, in particular, Eq. (\ref{DT-GR}) does apply in SET in the absence of gravitation, i.e., when $\beta \equiv 1$. Thus, if we do recover Eq. (\ref{divT_chg+field-GR}) of SR from SET in the absence of gravitation, then Eq. (\ref{divT_int=0}) will apply in that situation. Below, just to check that form of the interaction energy tensor that, we will show independently, is relevant, we therefore assume in advance that Eq. (\ref{divT_int=0}) applies to SET in the absence of gravitation, and we will show in Sect. \ref{ChargeConservn} that this is indeed the case with the precise framework that shall be adopted.

\vspace{3mm}
In view of the foregoing, it is not {\it a priori} obvious that we may impose the condition that the interaction tensor $\Mat{T}_{\mathrm{inter}}$ should vanish in SR. However, we may impose that it should be Lorentz-invariant in SR. Any Lorentz-invariant second-order tensor is a scalar multiple of the Minkowski metric tensor, say $\Mat{\gamma }^0$ \cite{A58}. Therefore, $\Mat{T}_{\mathrm{inter}}$ should have the form (in Cartesian coordinates for the Minkowski metric, i.e., $(\gamma^0)_{\mu \nu }=\eta _{\mu \nu } $):
\be\label{T_inter_SR}
T_{\mathrm{inter}\ \mu \nu }= p\, \eta_{\mu \nu }\qquad \mathrm{(SR)},
\ee
with some scalar field $p$. This defines of course a Lorentz-invariant tensor field. Note that (\ref{T_inter_SR}) is equivalent to:
\be\label{T_inter_SR_mixed}
T^\mu_{\mathrm{inter}\ \ \nu }:= \eta ^{\mu \rho } \,p\, \eta_{\rho  \nu } = p\,\delta ^\mu _\nu \qquad \mathrm{(SR)}
\ee
(again in Cartesian coordinates). Now we observe that the definition
\be\label{T_inter_mixed}
T^\mu_{\mathrm{inter}\ \ \nu }:=  p\,\delta ^\mu _\nu,
\ee
thus got in Cartesian coordinates in a Minkowski spacetime, is actually generally-covariant: if (\ref{T_inter_mixed}) applies in some coordinates $x^\mu$ in a general spacetime with metric $\Mat{\gamma }$, it still applies after any coordinate change. Therefore, we adopt for the general case the definition (\ref{T_inter_mixed}), which has been got by demanding that $\Mat{T}_{\mathrm{inter}}$ be Lorentz-invariant in SR. When gravitation is absent i.e. when the metric turns out to be Minkowski's, we should have Eq. (\ref{divT_int=0}), and this implies $p=\mathrm{Constant}$, say $p\equiv p_\infty $. Indeed, if the metric is Minkowski's, Eqs. (\ref{divT_int=0}) and (\ref{T_inter_SR}) imply that $p_{,\mu}=0$. Thus by requiring only that $\Mat{T}_{\mathrm{inter}}$ should be {\it Lorentz-invariant} in SR, we get that it is then actually {\it constant}. Moreover, it is natural to assume that, very far from any body, the total energy tensor $\Mat{T}$, as well as $\Mat{T}_{\mathrm{chg}}$ and $\Mat{T}_{\mathrm{field}}$, are zero. That assumption implies that the constant $p_\infty $ is zero, and hence that the additivity condition (\ref{T=Tcharges+Tfield}) applies in SR. We note also that, with the definition (\ref{T_inter_mixed}), we have just one unknown more ($p$) in the system of electrodynamics of SET [(\ref{M_lambda mu nu}), (\ref{Dyn_T-chg-SET}), (\ref{DT-SET})], see the end of Sect. \ref{SET_without_T3}. So we have to find just one scalar equation more. Thus we cannot add the standard Maxwell second group (\ref{Maxwell 2 GR}), since it would add four independent scalar equations. We have to show this more clearly, because it is an important point. \\

Once an interaction tensor is introduced through Eq. (\ref{T_with_interact}), we may {\it a priori} postulate the standard Maxwell second group (\ref{Maxwell 2 GR}) in addition to the system of electrodynamics of SET [(\ref{M_lambda mu nu}), (\ref{Dyn_T-chg-SET}), (\ref{DT-SET})] \cite{A56}. This leads to (Eq. (104) in Ref. \cite{A56}): 
\be\label{Dyn-T_Interact with MaxwellGR}
T_\mathrm{inter \ \,;\nu}^{\mu \nu} - b^\mu(\Mat{T}_\mathrm{inter}) = b^\mu(\Mat{T}_\mathrm{field}).
\ee
Let us compute the l.h.s. (this will be used also in the next section). Until the end of this section, and again in Sect. \ref{WeakField-p}, we will use coordinates $x^\mu$ that are adapted to the preferred reference f\mbox{}luid $\mathcal{E}$ and, more specifically, we will assume that the spatial coordinates are Cartesian for the Euclidean spatial metric $\Mat{g}^0$ assumed in the theory ($\Mat{g}^0$ is time-independent in any coordinates that are adapted to the preferred frame $\mathcal{E}$), and $x^0=cT$ with $T$ the preferred time of the theory. In such coordinates, the curved ``physical" spacetime metric $\Mat{\gamma }$ is by assumption \cite{A35}:
\be\label{gamma}
\dd s^2 = \gamma_{\mu \nu} \dd x^\mu \dd x^\nu = \beta^2 (\dd x^0)^2 - g_{ij} \dd x^i \,\dd x^j = \beta^2 (\dd x^0)^2 - \beta^{-2} \dd x^i \,\dd x^i.
\ee
This implies that, in such coordinates, we have 
\be\label{gamma_SET}
\gamma :=\mathrm{det}\,(\gamma _{\mu \nu })=-\beta ^{-4}. 
\ee
With the help of an identity for $T_{\mu\ \ ;\nu}^{\ \,\nu}$ (\cite{L&L}, Eq. (86.11)) and the definition (\ref{b^mu}) of $b^\mu$, we obtain thus (for whatever symmetric tensor $T_{\mu \nu }$):
\be\label{(divT-b)_0}
T_{ 0\ \ ;\nu}^{\ \nu} - b_0(\Mat{T}) \equiv \beta ^2\left ( T^{0 0}_{,0}+ T^{0 j}_{,j}\right ) -\beta \beta _{,0}\,T^{0 0},
\ee
\be\label{(divT-b)_i}
T_{ i\ \ ;\nu}^{\ \nu} - b_i(\Mat{T}) \equiv \beta ^2\left ( \frac{T_{ i}^{\ \nu}}{\beta ^2} \right )_{,\nu } -\beta^{-3} \beta _{,0}\,T^{0 i}-\beta^{-3} \beta _{,i}\,T^{j j}-\beta \beta _{,i}\,T^{0 0}.
\ee
With the definition just adopted (\ref{T_inter_mixed}) for $\Mat{T}_{\mathrm{inter}}$, we get from this: 
\be\label{(divT_int-b)_0}
\delta _0  = p_{,0} - 3\,p \,\beta _{,0}\,\beta^{-1},
\ee
\be\label{(divT_int-b)_i}
\delta _i  = p _{,i},
\ee
where
\be\label{Def_delta_mu} 
\delta _\mu := \delta _\mu(p) := T_{\mathrm{inter}\  \mu \ \ ;\nu}^{\ \nu} - b_\mu (\Mat{T}_\mathrm{inter}).
\ee
Equation (\ref{Dyn-T_Interact with MaxwellGR}) is equivalent to 
\be\label{delta=b}
\delta _\mu(p) = b_\mu(\Mat{T}_\mathrm{field})\quad (\mu =0,...,3). 
\ee
Clearly, for a general e.m. and gravitational field, $b_\mu(\Mat{T}_\mathrm{field})$ can be essentially any 4-vector field. (This can be checked by using the explicit forms of $\Mat{T}_\mathrm{field}$ and $b^\mu$, Eqs. (\ref{T field}) and (\ref{b^mu}).) Hence, Eqs. (\ref{(divT_int-b)_0}) and (\ref{(divT_int-b)_i}) imply that it cannot in general exist a scalar field $p$ that verify the four equations (\ref{delta=b}): already the three spatial equations $(\mu =i=1,2,3)$ imply that $\mathrm{rot}\,(b_i)={\bf 0}$, which is not true in general.\\

Thus, postulating the validity of the standard Maxwell second group (\ref{Maxwell 2 GR}) would demand to have a more general form than (\ref{T_inter_mixed}) for the interaction tensor $\Mat{T}_{\mathrm{inter}}$, which instead should depend on four scalar fields. In that case, the interaction tensor would not be Lorentz-invariant in SR any more. It is hard to say in advance whether or not it would then be possible to get the constancy of $\Mat{T}_{\mathrm{inter}}$ in SR --- which applies to the single-scalar form (\ref{T_inter_mixed}) as we will show in the next section. %Moreover, such more general form for $\Mat{T}_{\mathrm{inter}}$ would mean that this would be like the energy tensor of some additional material medium, which would be present for some mysterious reason each time that a charged medium is there with its e.m. field.

%%%%%%%%%%%%%%%%%%%%%%%%%%%%%%%%%%%%%%%%%%%%%%%%%%%%%%%%%%%%%%%%%%%%%%%%%%%%%%%%
\section{Charge conservation and determination of the interaction tensor}\label{ChargeConservn}
%%%%%%%%%%%%%%%%%%%%%%%%%%%%%%%%%%%%%%%%%%%%%%%%%%%%%%%%%%%%%%%%%%%%%%%%%%%%%%%%

With the general decomposition (\ref{T_with_interact}), the dynamical equation (\ref{DT-SET}) for the total energy tensor in SET is equivalent to:
\be\label{divT_chg+field+inter}
T_{\mathrm{field}\ \,;\nu}^{\mu  \nu} = b^\mu(\Mat{T}_\mathrm{field}) + b^\mu (\Mat{T}_\mathrm{chg})-T^{\mu  \nu}_{\mathrm{chg}\ \ ;\nu} +b^\mu(\Mat{T}_\mathrm{inter}) -T_\mathrm{inter \ \,;\nu}^{\mu \nu}.
\ee  
By using the identity (\ref{Eq T-field-2}) for $T_{\mathrm{field}\ \,;\nu}^{\mu  \nu}$, the dynamical equation (\ref{Dyn_T-chg-SET}) for the charged medium, and the definition (\ref{Def_delta_mu}), this rewrites as
\be\label{divT_chg+field+inter-2}
F^\mu_{\ \ \lambda }\,F^{ \lambda \nu }_{\ \ ;\nu } = \mu _0 \left[ b^\mu (\Mat{T}_\mathrm{field})-F^\mu_{\ \, \nu }\,J^\nu -\delta^\mu(p)\right ].
\ee
If the matrix $(F^\mu_{\ \ \lambda })$ is invertible, which is the generic situation and is equivalent to ${\bf E.B} \ne 0$ \cite{A56}, this can still be rewritten as 
\be\label{Maxwell 2 SET p}
F^{\mu  \nu }_{\ \ ;\nu } = \mu _0 \left[ G^\mu_{\ \ \nu  } \left (b^\nu (\Mat{T}_\mathrm{field})-\delta ^\nu (p)\right ) -J^\mu \right ],
\ee
where $(G^\mu_{\ \ \nu  })$ is the inverse matrix of matrix $(F^\mu_{\ \ \nu  })$. By using the identity (\ref{divdivF=0}), we get from this:
\be\label{divJ SET p}
J^\mu_{\ \, ;\mu} = \left[ G^\mu_{\ \ \nu  } \left (b^\nu (\Mat{T}_\mathrm{field})-\delta ^\nu (p)\right ) \right ]_{;\mu}.
\ee
Apart from the identities (\ref{Eq T-field-2}) and (\ref{divdivF=0}) (which are valid independently of any physical theory), the validity of Eqs. (\ref{Maxwell 2 SET p}) and (\ref{divJ SET p}) depends only on the validity of Eqs. (\ref{DT-SET}) and (\ref{Dyn_T-chg-SET}), accounting for the general decomposition (\ref{T_with_interact}) with the interaction energy tensor (\ref{T_inter_mixed}).  In Sect. \ref{Constraints}, we showed that, to close the system of electrodynamics of SET [(\ref{M_lambda mu nu}), (\ref{Dyn_T-chg-SET}), (\ref{DT-SET})] in the presence of the interaction energy tensor (\ref{T_inter_mixed}), we need just one scalar equation more. Therefore, it suggests itself to close the system by adding the conservation of charge $J^\mu_{\ \, ;\mu}=0$, i.e., in view of (\ref{divJ SET p}), by adding the following equation:
\be\label{conservJ SET p}
\left[ G^{\mu \nu  } \left (b_\nu (\Mat{T}_\mathrm{field})-\delta _\nu (p)\right ) \right ]_{;\mu} =0.
\ee
Thus we have the system [(\ref{M_lambda mu nu}), (\ref{Dyn_T-chg-SET}), (\ref{DT-SET}), (\ref{conservJ SET p})], or equivalently the system [(\ref{M_lambda mu nu}), (\ref{Dyn_T-chg-SET}), (\ref{Maxwell 2 SET p}), (\ref{conservJ SET p})], which has the differential identity (\ref{Das_GR}). So we have $13-1=12$ equations for 12 unknowns $F_{\mu \nu }\ (0\leq \mu<\nu\leq 3)$, ${\bf J}$, $\rho^\ast $, and $p$.  %this system seems just as well determined as is the system of electrodynamics of SR and 
It is natural to expect that the system [(\ref{M_lambda mu nu}), (\ref{Dyn_T-chg-SET}), (\ref{Maxwell 2 SET p}), (\ref{conservJ SET p})] has a unique solution when suitable boundary conditions are imposed, just as has the system of electrodynamics of a metric theory, [(\ref{Maxwell 2 GR})--(\ref{M_lambda mu nu}), (\ref{Dyn_T-chg-GR})]. (As at the end of Sect. \ref{SET_without_T3}, the fact that the scalar gravitational field $\beta $ is also unknown can be accounted for by adding the scalar f\mbox{}lat-spacetime wave equation for $\psi :=-\mathrm{Log}\,\beta $.)\\

Now consider the case of a constant gravitational field, which includes the situation without any gravitational field. In that case, the general dynamical equation of GR (\ref{DT-GR}) applies to SET as noted after Eq. (\ref{divT_int=0}), so that we get from the general decomposition (\ref{T_with_interact}) and the definition (\ref{T_inter_mixed}) of the interaction tensor:
\be\label{divT_chg+field+inter-const}
T_{\,;\nu}^{\mu  \nu}=0 = T_{\mathrm{field}\ \,;\nu}^{\mu  \nu} + T^{\mu  \nu}_{\mathrm{chg}\ \ ;\nu}  + T_\mathrm{inter \ \,;\nu}^{\mu \nu} =  T_{\mathrm{field}\ \,;\nu}^{\mu  \nu} + T^{\mu  \nu}_{\mathrm{chg}\ \ ;\nu}  + p_{,\nu} \gamma ^{\mu \nu }.
\ee 
Hence, the system [(\ref{M_lambda mu nu}), (\ref{Dyn_T-chg-SET}), (\ref{Maxwell 2 SET p}), (\ref{conservJ SET p})] is solved by $p=\mathrm{Constant}:=p_\infty $ and with the fields $\Mat{F}$, ${\bf J}$, $\rho^\ast $ being the solution of the system [(\ref{Maxwell 2 GR})--(\ref{M_lambda mu nu}), (\ref{Dyn_T-chg-GR})] for the given time-independent metric and for the relevant boundary conditions.  Therefore, we get that Eq. (\ref{divT_int=0}) is true in a constant gravitational field. In particular, Eq. (\ref{divT_int=0}) is true when the latter vanishes --- as we provisionally assumed after postulating the form (\ref{T_inter_mixed}) of the interaction energy tensor, to check that form. Moreover, as noted after Eq. (\ref{T_inter_mixed}), we may assume that the constant $p_\infty $ is zero, hence the additivity  (\ref{T=Tcharges+Tfield}) of the energy tensors $\Mat{T}_\mathrm{chg}$ and $ \Mat{T}_\mathrm{field} $ does apply in a constant gravitational field.\\

If the e.m. field $\Mat{F}$ and the gravitational field $\beta $ are considered given, then the scalar field $p$, and hence the interaction energy tensor, may in principle be calculated by solving the equation for charge conservation, Eq. (\ref{conservJ SET p}). Of course $\Mat{F}$ and $\beta $ are coupled with the other fields that include precisely $p$. %but, due to the foregoing, we expect that, for the (usual) case of a slowly variable gravitational field, $\Mat{F}$ and $\beta $ are close to  

%%%%%%%%%%%%%%%%%%%%%%%%%%%%%%%%%%%%%%%%%%%%%%%%%%%%%%%%%%%%%%%%%%%%%%%%%%%%%%%%
\section{First approximation of the scalar field in a weak gravitational field}\label{WeakField-p}
%%%%%%%%%%%%%%%%%%%%%%%%%%%%%%%%%%%%%%%%%%%%%%%%%%%%%%%%%%%%%%%%%%%%%%%%%%%%%%%%

In this section we will obtain equations that should allow one, in a future work, to assess numerically the scalar field $p$ and the interaction energy in a given e.m. field and in a given {\it weak and slowly varying} gravitational field. To this aim, we may use the results of the asymptotic framework developed in some detail in Sects. 4 and 5 of Ref. \cite{A56}. The only change with respect to that former work is the additional term $-G^{\mu \nu  }\delta _\nu (p)$ in Eq. (\ref{Maxwell 2 SET p}), as compared with Eq. (22) in Ref. \cite{A56}, and correspondingly the additional equation (\ref{conservJ SET p}). Therefore, the analysis done in that work applies almost without any change. In view of the additional equation (\ref{conservJ SET p}), we must require that $\delta _\nu (p)$ be of the same order as is $b_\nu (\Mat{T}_\mathrm{field})$ in the gravitational weak-field parameter $\lambda $, with $\lambda =c^{-2}$ in specific $\lambda $-dependent units of mass and time. From Eqs. (34) and (36) in Ref. \cite{A56}, we have for the order of $b_\nu $:
\be
b_\nu (\Mat{T}_\mathrm{field}) =\mathrm{ord}(c^{2n-5}), 
\ee
where 
\be
\Mat{F} = c^n \left(\overset{0}{\Mat{F}} + c^{-2}\,\overset{1}{\Mat{F}} +O(c^{-4})\right) 
\ee
is the expansion of the e.m. field tensor. The gravitational field is expanded as (Eq. (28) in Ref. \cite{A56}):  
\be\label{1-U/c2}
\beta := \sqrt{\gamma _{00}} = 1 - U\,c^{-2} + O(c^{-4}),
\ee
where $U$ is the Newtonian gravitational potential. Using this and setting 
\be\label{expans p}
p = c^q \left(\overset{0}{p} + c^{-2}\,\overset{1}{p} +O(c^{-4})\right),
\ee
we get from (\ref{(divT_int-b)_0}) and (\ref{(divT_int-b)_i}) that $\delta _{\mu}=\mathrm{ord}(c^q)$, hence our requirement is satisfied iff $q = 2n-5$ so that
\be\label{expans p_,mu}
p_{,\mu } = c^{2n-5} \left(\overset{0}{p}_{,\mu }  +O(c^{-2})\right).
\ee
Let us define
\be\label{Charge rate ETG}
\hat{\rho  } :=  \left( G^\mu  _{\ \,\nu }\,b^\nu (\Mat{T}_\mathrm{field}) \right )_{; \mu} ,
\ee
as in Eq. (23) in Ref. \cite{A56} --- but now $\hat{\rho  }\ne  J^\mu  _{\ \,; \mu}$, unlike in the latter work. With this definition, everything in Sects. 4 and 5 of Ref. \cite{A56} remains valid and we have in particular (Eq. (45) in Ref. \cite{A56}):
\footnote{\
The remainder in this expansion was incorrectly written as $O\left(c^{-5}\right)$ in Eqs. (45) and (54) of Ref. \cite{A56}. In fact it is $O\left(c^{n-7}\right)$ while the main term is $\mathrm{ord}\left(c^{n-5}\right)$, Eq. (41) of Ref. \cite{A56}. The important point is precisely that the remainder is $O\left(c^{-2}\right)$ times the main term.
}
\be\label{expans-rho-hat-2}
\hat{\rho } = c^{-3}  \left[ \left( G_1\,^{\mu  0}\, T_1\,^{j j}- G_1\,^{\mu i}\, T_1\,^{0 i}\right) \partial _T U \right]_{,\mu }\left (1+O\left(c^{-2}\right)\right),
\ee
where $G_1\,^{\mu  \nu}$ and $T_1\,^{\mu  \nu}$ are the first approximations of $G\,^{\mu  \nu}$ and $T\,^{\mu  \nu}$, i.e., 
\be\label{G_1 & F_1}
\Mat{G}_1 := (\Mat{F}_1)^{-1}, \qquad \Mat{F}_1 := c^{n}\overset{0}{\Mat{F}},
\ee
and the like for $\Mat{T}_1$ \cite{A56}. This can be then calculated explicitly: as given by Eqs. (53) and (54) of Ref. \cite{A56},
\be\label{expans-rho-hat-3}
\hat{\rho } = c^{-3}  \left( e^i \partial _T U \right)_{,i } \left (1+O\left(c^{-2}\right)\right),
\ee
where $e^i =\frac{1}{2\, c\, \mu_0\, \left(B_1\, E_1 + B_2\, E_2 + B_3\, E_3\right)}\times $
\be\label{e^i}
 \times \left(\begin{array}{c} c^2 \left ({B_1}^3  + B_1\, {B_2}^2 + B_1\, {B_3}^2 \right ) + B_1\, {E_1}^2 - B_1\, {E_2}^2 - B_1\, {E_3}^2 + 2\, B_2\, E_1\, E_2 + 2\, B_3\, E_1\, E_3\\
c^2 \left ({B_2}^3  + B_2\, {B_3}^2 + B_2\, {B_1}^2 \right ) + B_2\, {E_2}^2 - B_2\, {E_3}^2 - B_2\, {E_1}^2 + 2\, B_3\, E_2\, E_3 + 2\, B_1\, E_2\, E_1\\ 
c^2 \left ({B_3}^3  + B_3\, {B_1}^2 + B_3\, {B_2}^2 \right ) + B_3\, {E_3}^2 - B_3\, {E_1}^2 - B_3\, {E_2}^2 + 2\, B_1\, E_3\, E_1 + 2\, B_2\, E_3\, E_2\end{array}\right).
\ee
%As precised in Ref. \cite{A56}, Eqs. (\ref{expans-rho-hat-3})--(\ref{e^i}) are valid in spatial coordinates that are adapted to the frame $\mathcal{E}$ and, more specifically, are Cartesian for the Euclidean spatial metric $\Mat{g}^0$; and $T$ is the preferred time of the theory. 
In Eq. (\ref{e^i}), $E_i:=E^i$ and $B_i:=B^i$ are the components of the first approximations of the electric and magnetic fields in the frame $\mathcal{E}$, in coordinates of the class specified after Eq. (\ref{Dyn-T_Interact with MaxwellGR}). I.e., $E^i$ and $B^i$ are extracted (Note \ref{F vs E-B}) from the first approximation $\Mat{F}_1$ of the e.m. field tensor $\Mat{F}$, Eq. (\ref{G_1 & F_1}), that obeys the f\mbox{}lat-spacetime Maxwell equations \cite{A56}. Inserting (\ref{expans p_,mu}) and (\ref{expans-rho-hat-3}) into (\ref{conservJ SET p}) using $u^\mu_{;\mu}=(u^\mu \sqrt{-\gamma })_{,\mu}/\sqrt{-\gamma }$, with $\sqrt{-\gamma }=1+O(c^{-2}) $ owing to (\ref{gamma_SET}) and (\ref{1-U/c2}), gives us:
\be\label{Eq for p}
\left (G_1^{\ \, \mu \nu }\, (p_1)_{,\nu}\right )_{,\mu}=G_{1\quad ,\mu}^{\ \, \mu \nu }\, (p_1)_{,\nu}= c^{-3} \left( e^i \partial _T U \right)_{,i }\left (1+O\left(c^{-2}\right)\right),
\ee
where $p_1 := c^{2n-5}\, \overset{0}{p}$ is the first approximation of $p$. The first equality is due to the antisymmetry of $G^{\mu \nu }$ and $G_1^{\ \, \mu \nu }$. The matrix $\Mat{G}':=(G_1^{\ \, \mu \nu })$ is given explicitly by Eq. (50) in Ref. \cite{A56}, which can be rewritten as
\be\label{G'}
\Mat{G}' = \frac{-c}{{\bf E.B}}\, \Mat{H},\quad \Mat{H}:= \left(\begin{array}{cccc} 0 & B_1 & B_2 & B_3\\ - B_1 & 0 & -\frac{E_3}{c} & \frac{E_2}{c}\\ - B_2 & \frac{E_3}{c} & 0 & -\frac{E_1}{c}\\ - B_3 & -\frac{E_2}{c} & \frac{E_1}{c} & 0 \end{array}\right).
\ee
It is easy to check that Maxwell's (f\mbox{}lat-spacetime) first group, verified by $\Mat{F}_1$, implies that 
\be
H^{\mu \nu }_{,\mu} = 0.
\ee
It follows from this and (\ref{G'}) that we have explicitly in (\ref{Eq for p}):
\be\label{k^nu}
(k^\nu) := \left(G_{1\quad ,\mu}^{\ \, \mu \nu }\right) =
\ee

$\frac{1}{{\left({\bf E.B}\right)}^2} \times \left(\begin{array}{c} - B_i\, c\, \left({\bf E.B}\right)_{,i}\\

E_3\, \left({\bf E.B}\right)_{,2}
- E_2\, \left({\bf E.B}\right)_{,3} 
+ B_1\, c\, \left({\bf E.B}\right)_{,0}\\

E_1\, \left({\bf E.B}\right)_{,3}
- E_3\, \left({\bf E.B}\right)_{,1} 
+ B_2\, c\, \left({\bf E.B}\right)_{,0}\\

E_2\, \left({\bf E.B}\right)_{,1}
- E_1\, \left({\bf E.B}\right)_{,2} 
+ B_3\, c\, \left({\bf E.B}\right)_{,0}
\end{array}\right)$,\\

i.e.,
\bea
k^0 & = & \frac{-c}{{\left({\bf E.B}\right)}^2}\, {\bf B.(\nabla ({\bf E.B}))},\\
(k^i) & = &\frac{1}{{\left({\bf E.B}\right)}^2}\, \left(\frac{\partial \left( {\bf E.B}\right)}{\partial T} {\bf B}-{\bf E}\wedge (\nabla ({\bf E.B})) \right ).
\eea

\vspace{2mm}
Equation (\ref{Eq for p}) can be rewritten in the form
\be\label{advec_p}
\partial _T\, p_1 + u^j \partial _j\,p_1 = S,
\ee
where
\be
S := \frac{c^{-2}\left( e^i \partial _T U \right)_{,i }}{k^0}
\ee
(no confusion can occur with the sum $S$ in Subsect. \ref{J given}), and
\be\label{Def u}
u^j := \frac{c\,k^j}{k^0}.
\ee
We assume $k^0 \ne 0$ in Eq. (\ref{k^nu}), i.e. ${\bf B.(\nabla ({\bf E.B}))} \ne 0$. Note that ${\bf E.B} \ne 0$ is required since Sect. \ref{ChargeConservn}. Remind that here the first-approximation fields ${\bf E}$ and ${\bf B}$ are involved, and they obey the f\mbox{}lat-spacetime Maxwell equations. Equation (\ref{advec_p}) is an advection equation with a given source $S$ for the unknown field $p_1$. This is a hyperbolic PDE whose characteristic curves are the integral curves of the  vector field ${\bf u} := (u^j)$. That is, on the curve $\mathcal{C}(T_0, {\bf x}_0)$ defined by
\be\label{cara}
\frac{\dd {\bf x}}{\dd T} ={\bf u}(T,{\bf x}), \qquad {\bf x}(T_0) = {\bf x}_0,
\ee
we have from (\ref{advec_p}):
\be\label{dp/dt}
\frac{\dd p_1}{\dd T} = \frac{\partial p_1}{\partial  T} + \frac{\partial p_1}{\partial  x^j} \frac{\dd x^j}{\dd T} = S(T,{\bf x}).
\ee
We note that the field ${\bf u}$ is {\it given,} Eq. (\ref{Def u}), i.e., it does not depend on the unknown field $p_1$. Therefore, the integral lines (\ref{cara}) are given, too, hence the characteristic curves do not cross. Thus, the solution $p_1$ is got uniquely by integrating (\ref{dp/dt}):
\be\label{p on C}
p_1(T,{\bf x}(T)) - p_1(T_0,{\bf x}_0) =\int _{T_0} ^T  S(t,{\bf x}(t))\, \dd t , 
\ee
where $T \mapsto {\bf x}(T)$ is the solution of (\ref{cara}). If at time $T_0$ the position ${\bf x}_0$ in the frame $\mathcal{E}$ is enough distant from material bodies, one may assume that $p_1(T_0,{\bf x}_0)=0$.

%%%%%%%%%%%%%%%%%%%%%%%%%%%%%%%%%%%%%%%%%%%%%%%%%%%%%%%%%%%%%%%%%%%%%%%%%%%%%%%%
\section{Conclusion}
%%%%%%%%%%%%%%%%%%%%%%%%%%%%%%%%%%%%%%%%%%%%%%%%%%%%%%%%%%%%%%%%%%%%%%%%%%%%%%%%

The main results obtained in this paper are the following ones:\\

1) The structure of classical electrodynamics based on the standard Maxwell equations of special relativity or general relativity has been discussed and it has been shown by using the notion of differential identity that the number of independent scalar PDE's is the same as the number of unknown fields. This applies to both the case with given 4-current and the more general case where one takes into account the equation of motion of the charged medium, the 4-current then belonging to the unknowns. Of course the result for the first case is well known (e.g. \cite{Jiang1996, Zhou2006}), but the explanation by considering the differential identities (\ref{Das_GR}) and (\ref{div_F+J=0}) is more straightforward and we could not find it in the literature. For the more general case also, our discussion is based on differential identities: (\ref{Das_GR}) is still valid but (\ref{div_F+J=0}) holds now only on the solution space; that discussion too seems new.\\

2) In the investigated theory of gravity (``SET"), with the additivity assumption (\ref{T=Tcharges+Tfield}), electrodynamics consists of the system [(\ref{M_lambda mu nu}), (\ref{Dyn_T-chg-SET}), (\ref{DT-SET})], i.e., Maxwell's first group and the dynamical equations for the charged medium and for the total energy tensor. Using the same method as for SR and GR, it has been shown that this is also a closed system of PDE's. While introducing the ``interaction energy tensor" $\Mat{T}_\mathrm{inter}$ by switching to (\ref{T_with_interact}), one necessarily introduces new unknowns, so that the foregoing system is not closed any more and one needs new equations. We imposed on $\Mat{T}_\mathrm{inter}$ that, in SR, it should be a Lorentz-invariant tensor. This determines that in the general case it has the form $T^\mu_{\mathrm{inter}\ \ \nu }:=  p\,\delta ^\mu _\nu$, with $p$ a scalar field (which, we showed, is constant and even zero in SR). Thus, only one additional equation is needed and this can consistently be imposed to be the charge conservation.\\

3) Considering a weak and slowly varying gravitational field, we derived the equation that determines the scalar field $p$, whose knowledge is equivalent to that of the interaction energy tensor: Eq. (\ref{advec_p}). We indicated how one may in principle compute that field in a given EM field and in a given  weak and slowly varying gravitational field, Eq. (\ref{p on C}). The interaction energy is gravitationally active, because its density $T_{\mathrm{inter}}^{0 0}= p\,\gamma ^{0 0}$ contributes to the total energy density of matter and non-gravitational fields, $T^{0 0}$. According to SET \cite{A35}, $T^{0 0}$ is the source of the gravitational field.  The interaction energy is not especially localized inside matter, and it has to be present in space as soon as there is matter that is electromagnetically active. It could thus be counted as ``dark matter". To learn more, it will be necessary to have recourse to a numerical work.\\

\noi {\bf Acknowledgement:} I am grateful to Jerzy Kijowski for a discussion on the motion of the sources. Also, a referee asked for more explanation about why the interaction energy could contribute to dark matter.

%%%%%%%%%%%%%%%%%%%%%%%%%%%%%%%%%%%%%%%%%%%%%%%%%%%%%%%%%%%%%%%%%%%%%%%%%%%%%%%%


\begin{thebibliography}{9}
\small

\bibitem{L&L}
Landau L.D., Lifshitz E.M., The classical theory of fields, 3rd English edition, Pergamon, Oxford, U.K., 1971 

\bibitem{Jackson1998}
Jackson J.D., Classical electrodynamics (3rd edn), Wiley, Hoboken (N.J.), U.S.A., 1998

\bibitem{A54} 
Arminjon M., Continuum dynamics and the electromagnetic field in the scalar ether theory of gravitation, Open Physics, 2016, 14, 395--409.

\bibitem{A56} 
Arminjon M., Charge conservation in a gravitational field in the scalar ether theory, Open Physics, 2017, 15, 877-890.

\bibitem{A35}
Arminjon M., Space isotropy and weak equivalence principle in a scalar theory of gravity, Braz. J. Phys., 2006, 36, 177--189. 

\bibitem{MTW} 
Misner C.W., Thorne K.S., Wheeler J.A.,  Gravitation, Freeman, San Francisco, U.S.A., 1973

\bibitem{Jiang1996} 
Jiang B.N., Wu J., Povinelli L.A., The origin of spurious solutions in computational electromagnetics, J. Comput. Phys., 1996, 125, 104-123.

\bibitem{Zhou2006} 
Zhou X.L.,  On independence, completeness of Maxwell's equations and uniqueness theorems in electromagnetics, Prog. Electromagn. Res., 2006, 64, 117-134.

\bibitem{Liu2017} 
Liu C., Explanation on overdetermination of Maxwell's equations, Physics and Engineering (in Chinese), 2017, 27(3), 7-9, and Preprint arXiv:1002.0892v9 (2018).

\bibitem{Das1996} 
Das A., The special theory of relativity -- a mathematical exposition, 2nd printing, Springer-Verlag, Berlin -- Heidelberg -- New York, 1996

\bibitem{Cattaneo1958}
Cattaneo C., General relativity: relative standard mass, momentum, energy and gravitational field in a general system of reference, Nuovo Cim., 1958, 10, 318-337.

\bibitem{B39} 
Arminjon M., On continuum dynamics and the electromagnetic field in the scalar ether theory of gravitation, J. Phys. Conf. Ser., 2017, 845, 012014 (9 pages).

\bibitem{Fock1964}
Fock V.A., The theory of space, time and gravitation, 2nd English edition, Pergamon, Oxford, U.K., 1964

\bibitem{Chandrasekhar1969}
Chandrasekhar S., Conservation laws in general relativity and in the post-Newtonian approximations, Astrophys. J., 1969, 158, 45-54.

\bibitem{A16}
Arminjon M., On the extension of Newton's second law to theories of gravitation in curved space-time, Arch. Mech., 1996, 48, 551--576.

\bibitem{A20}
Arminjon M., On the possibility of matter creation/destruction in a variable gravitational field, Analele Universit. Bucure\c{s}ti -- Fizic\u{a}, 1998, 47, 3-21.

\bibitem{Moller1952}
M\o ller C., The theory of relativity, Clarendon Press, Oxford, U.K., 1952

\bibitem{Muller1968}
M\"uller I., A thermodynamic theory of mixtures of f\mbox{}luids, Arch. Rational Mech. Anal., 1968, 28, 1--39.

\bibitem{A58}
Arminjon M., Lorentz-invariant second-order tensors and an irreducible set of matrices, submitted for publication, and preprint \href{https://hal.archives-ouvertes.fr/view/index/docid/1797592}{HAL-01797592}, 2018.

\end{thebibliography}
\end{document}